\documentclass{article}
\usepackage[cmfonts]{eusipco2009}
\usepackage{amsmath,epsfig,cite}

\def\vec{\mathrm{vec}}

\def\argmin{\mathrm{argmin}}

\def\bA{\mathbf{A}}
\def\bE{\mathbf{E}}
\def\bG{\mathbf{G}}
\def\bI{\mathbf{I}}
\def\bR{\mathbf{R}}
\def\bW{\mathbf{W}}

\def\btheta{\mbox{\boldmath{$\theta$}}}
\def\bsigma{\mbox{\boldmath{$\sigma$}}}
\def\bSigma{\mbox{\boldmath{$\Sigma$}}}

\title{Self-Calibration of Radio Astronomical Arrays With Non-Diagonal
  Noise Covariance Matrix}

\name{Stefan J. Wijnholds and Alle-Jan van der Veen \thanks{This work was
    supported by the Netherlands Institute for Radio Astronomy (ASTRON) and by
  NWO-STW under the VICI programme (DTC.5893).}}

\address{
\begin{minipage}[t]{.45\textwidth}
\begin{center}
ASTRON\\
R\&D Department\\
Oude Hoogeveensedijk 4, NL-7991 PD, Dwingeloo, The Netherlands\\
phone: +31 521 595 261, email: wijnholds@astron.nl\\
web: www.astron.nl
\end{center}
\end{minipage}
\begin{minipage}[t]{0.05\textwidth}
\end{minipage}
\begin{minipage}[t]{.45\textwidth}
\begin{center}
Delft University of Technology\\
Department of Electrical Engineering\\
Mekelweg 4, NL-2628 CD, Delft, The Netherlands\\
phone: +31 15 278 6240, email: allejan@cas.et.tudelft.nl\\
web: www.tudelft.nl
\end{center}
\end{minipage}
}

\begin{document}

\maketitle

\begin{abstract}
  The radio astronomy community is currently building a number of phased array
  telescopes. The calibration of these telescopes is hampered by the fact that
  covariances of signals from closely spaced antennas are sensitive to noise
  coupling and to variations in sky brightness on large spatial scales. These
  effects are difficult and computationally expensive to model. We propose to
  model them phenomenologically using a non-diagonal noise covariance
  matrix. The parameters can be estimated using a weighted alternating least
  squares (WALS) algorithm iterating between the calibration parameters and
  the additive nuisance parameters. We demonstrate the effectiveness of our
  method using data from the low frequency array (LOFAR) prototype station.
\end{abstract}


\section{Introduction}
\label{sec:intro}

The radio astronomical community is currently constructing a number of large
scale phased array telescopes such as the low-frequency array (LOFAR)
\cite{Vos2009-1} and the Murchison wide field array (MWA)
\cite{Lonsdale2008-1}. These instruments need to be calibrated regularly to
track variations in the electronics of the antennas and receivers, as well as
direction dependent variations of the ionosphere \cite{Tol2007-1}. E.g., LOFAR
will consist of order 50 stations (distributed over an area of a hundred
kilometers or more), where each station consists of 96 ``low band''
dual-polarized dipole antennas (10--90 MHz) and 96 ``high band'' antennas
(110--240 MHz). The latter antennas are in turn composed of 16 beamformed
dual-polarized droopy dipole antennas. Each station provides a number of
beamformed outputs, which in turn are correlated at a central location to form
images and other astronomy products.

In this paper we focus on the calibration of the station antennas. The general
aim is to estimate the direction independent gains and phases of each sensor,
as well as the direction dependent gains corresponding to each source.  This
is done for the 2--10 brightest sources in the sky, assuming a point source
model.  The problem is complicated by the fact that the covariances of signals
from closely spaced antennas within a station are sensitive to noise coupling
(for the lowest frequencies, the antennas are spaced closer than half a
wavelength). Also, the point source model does not entirely hold because of
bright emission from the plane of the galaxy, extending over the entire
sky. Fortunately, this emission is spatially smooth, which implies that it is
dominant on the short spatial scales in the array aperture, i.e. on the short
baselines. In this paper we propose to model both short baseline effects by an
additive noise covariance matrix, which in this case is not diagonal and has
unknown entries for each short baseline.  If we can estimate this matrix, a
simple point source model will be sufficient to calibrate the array, which
reduces the problem to a problem for which solutions are readily available
\cite{Fuhrmann1994-1, Wijnholds2006-1, Wijnholds2008-1}.

Direction finding problems for calibrated arrays in the presence of unknown
correlated noise have been extensively studied in the 1990s.  It was proven
that the general problem is not tractable without imposing some appropriate
constraints on the noise covariance matrix or exploiting differences in
temporal characteristics between source and noise signals
\cite{Stoica1992-1}. Radio astronomical signals generally behave like noise,
thus temporal techniques (instrumental variables) are not applicable. Instead,
we should rely on an appropriately constrained parameterized model of the
noise covariance matrix. Starting with \cite{bohme88icassp}, a series of
papers were published; see \cite{Goransson1999-1} for an overview.  ML
estimators for the source and instrument parameters under a generalized noise
covariance parameterization is provided in \cite{friedlander95tsp,
  Goransson1999-1}, whereas nonlinear least squares estimators were studied in
\cite{friedlander95tsp, Wax1996-1, Ottersten1998-1}. In either case, an
analytic source and instrument parameter dependent solution is derived for the
noise model parameters which is substituted back into the cost function. This
cost function then has to be minimized using a generalized solving technique,
such as Newton iterations. This approach works well if the number of
instrument and source parameters is small. For larger problems (we consider
100 antenna/source parameters and over 750 noise covariance parameters), it is
convenient to exploit suboptimal but closed-form analytic solutions, at least
for initialization.  We therefore propose a weighted alternating least squares
(WALS) approach which iterates over noise, source and instrument
parameters. The proposed method can thus be regarded as an extension to the
methods proposed in \cite{Wijnholds2008-1}

\emph{Notation}: The transpose operator is denoted by $^T$, the complex
conjugate (Hermitian) transpose by $^H$, complex conjugation by
$\overline{(\cdot)}$ and the pseudo-inverse by $^\dagger$. An estimated value
is denoted by $\widehat{(\cdot)}$. $\otimes$ denotes the Kronecker product and
$\circ$ is used to denote the Khatri-Rao or column-wise Kronecker product of
two matrices. $\vec(\cdot)$ converts a matrix to a vector by stacking the
columns of the matrix.

\section{Data Model and problem statement}
\label{sec:model}

We consider an array of $P$ antennas. The measured $P \times P$ array
covariance matrix can be modeled as
\begin{equation}
\bR = \bR_0 \left ( \btheta \right ) + \bSigma_n
\end{equation}
where $\bR_0 \left ( \btheta \right )$ is the signal model for an ideal noise
free array, which depends on a number of unknown real valued parameters
accumulated in a column vector $\btheta$, and $\bSigma_n$ is a $P \times P$
matrix describing the noise corruption.  Note that $\bSigma_n$ must be
Hermitian, since the array covariance matrix is a Hermitian matrix.  In our
application, the data covariance model is
\begin{equation}\label{eq:noisefree}
   \bR_0 \left ( \btheta \right ) = \bG_1 \bA \bG_2 \bSigma_s \bG_2^H \bA^H
   \bG_1^H
\end{equation}
where $\bSigma_s$ is the covariance of the point sources (assumed to be known
from tables), $\bA$ contains the direction vectors, $\bG_1$ a (diagonal)
instrumental gain/phase matrix, and $\bG_2$ a (diagonal) direction dependent
gain matrix \cite{Wijnholds2008-1}. If the direction dependent gains are
unknown, we can introduce $\bSigma = \bG_2 \bSigma_s \bG_2^H$, which implies
that we should estimate the apparent source powers. The direction dependent
gains follow directly from the apparent source powers if the actual source
powers are known, e.g.\ from tables. If the source positions are unknown or
perturbed by propagation conditions, $\bA$ may be parameterized
\cite{Wijnholds2008-1}. The contents of the parameter vector $\btheta$
therefore strongly depend on the available knowledge on the instrument, the
propagation conditions and the sources.

As in \cite{bohme88icassp} and subsequent papers, the unknown noise covariance
matrix is modeled as a linear sum of known matrices, in this case simple
selection matrices $\bE_{ij}$ which are zero everywhere except for a '1' in
entry $(i,j)$,
\[
   \bSigma_n = \sum_{(i,j) \in \cal{S}}  \sigma_{ij} \bE_{ij} \,.
\]
The set $\cal{S}$ contains the index pairs of the short baselines, including
the autocorrelation entries $(i,i)$.  The unknown coefficients $\sigma_{ij}$
are the nuisance parameters.

In the absence of the noise corruptions, i.e., $\bR = \bR_0 \left ( \btheta
\right )$, the estimation problem to find the parameter vector $\btheta$ is
commonly formulated either as a ML problem, or as a generalized least squares
estimation problem
\begin{equation}
  \widehat{\btheta} = \underset{\btheta}{\argmin} \left \Arrowvert \bW
    \left ( \widehat{\bR} - \bR_0 \left ( \btheta \right ) \right ) \bW \right
  \Arrowvert_F^2, \label{eq:WLS_theta}
\end{equation}
where $\widehat{\bR}$ is the measured array covariance matrix. It is known
that with $\bW = \bR^{-1/2}$ this estimator is asymptotically unbiased and
asymptotically efficient \cite{friedlander95tsp, Ottersten1998-1}. Indeed, the
simulations in \cite{friedlander95tsp} show only a small improvement of the ML
solution as compared to the WLS solution. For $\bR_0(\btheta)$ given in
(\ref{eq:noisefree}), this problem, as well as the case with an unknown
diagonal noise covariance, was studied by us in \cite{Wijnholds2008-1}, in the
present paper, we will assume that a solution to this problem is available.

In the presence of correlated noise on the short baselines, the problem 
is extended to
\begin{equation}
\left \{ \widehat{\btheta}, \widehat{\bsigma}_n \right \} =
\underset{\btheta,\bsigma_n}{\argmin} \left \Arrowvert \bW \left (
    \widehat{\bR} - \bR_0 \left ( \btheta \right ) - \bSigma_n \right ) \bW
\right \Arrowvert_F^2,
\label{eq:problem}
\end{equation}
where $\bsigma_n$ is a vector containing all unique real valued parameters
required to describe the nonzero entries of $\bSigma_n$. This vector can be
related to $\bSigma_n$ using a selection matrix $\bI_s$ such that $\vec \left (
  \bSigma_n \right ) = \bI_s \bsigma_n$. By choosing the selection matrix
appropriately, we can ensure that the estimated $\bSigma_n$ is Hermitian.

\section{Parameter estimation}
\label{sec:wals}

\subsection{Weighted Alternating Least Squares}

We propose to solve the problem in Eq.\ \eqref{eq:problem} by alternating
between weighted least squares (WLS) estimation of the desired parameters
$\btheta$ and WLS estimation of the nuisance parameters $\bsigma_n$. The first
WLS problem can be formulated as
\begin{equation}
\widehat{\btheta} = \underset{\btheta}{\argmin} \left \Arrowvert \bW
  \left ( \left ( \widehat{\bR} - \bSigma_n \right ) - \bR_0 \left ( \btheta
    \right ) \right ) \bW \right \Arrowvert_F^2. \label{eq:WLS_theta2}
\end{equation}
which is identical in form to Eq. \eqref{eq:WLS_theta} for which a solution is
assumed to be available.

The second WLS problem can be formulated as
\begin{eqnarray}
\widehat{\bsigma}_n & = & \underset{\bsigma_n}{\argmin} \left \Arrowvert \bW
  \left ( \left ( \widehat{\bR} - \bR_0 \left ( \btheta \right ) \right ) -
    \bSigma_n \right ) \bW \right \Arrowvert_F^2 \nonumber\\
& = & \underset{\bsigma_n}{\argmin} \Big \Arrowvert \left ( \overline{\bW}
    \otimes \bW \right ) \vec \left ( \widehat{\bR} - \bR_0 \left ( \btheta
    \right ) \right ) - \nonumber\\
& & \left (\overline{\bW} \otimes \bW \right ) \bI_s \bsigma_n \Big
\Arrowvert_F^2.
\end{eqnarray}
The solution to this problem is given by
\begin{equation}
\widehat{\bsigma}_n = \left ( \left ( \overline{\bW} \otimes \bW \right )
  \bI_s \right )^\dagger \left ( \overline{\bW} \otimes \bW \right ) \vec
\left ( \widehat{\bR} - \bR_0 \left ( \btheta \right ) \right
).\label{eq:WLS_nuisance}
\end{equation}

In section \ref{sec:model} it was mentioned that $\bW = \bR^{-1/2}$ provides
optimal weighting for the LS cost function. Since $\bR$ is not known,
$\widehat{\bR}$ is used instead in many applications. It can be shown that
this may lead to a bias in the estimate of $\widehat{\bsigma}_n$ for a finite
number of samples \cite{Wijnholds2008-1}. This bias can be avoided by using
the best available model $\bR \left ( \widehat{\btheta}, \widehat{\bsigma}_n
\right )$ instead of $\widehat{\bR}$.

Estimation of receiver noise powers is a special case of the general problem
treated here. In this case $\bSigma_n$ is a diagonal, and $\bI_s = \bI \circ
\bI$ where $\bI$ is the $P \times P$ identity matrix. This form of selection
matrix simplifies Eq.\ \eqref{eq:WLS_nuisance} considerably
\cite{Wijnholds2008-1}. In some problems, for example in estimating the
receiver based gains, it is then possible to simply ignore the diagonal
entries instead of including nuisance parameters \cite{Wijnholds2006-1}. It
can further be shown that ignoring the corrupted entries instead of including
them using nuisance parameters does not change the Cram\`er-Rao bound of the
parameters of interest \cite{Tol2006-1}. This can be explained intuitively by
regarding the matrix equation describing the WLS problem as a set of scalar
equations. If a unique nuisance parameter is added to one of those scalar
equations, that equation is required to solve for the nuisance parameter and
can thus not be used to solve any other parameters. This implies that this
equation could have been ignored if one would only focus on the parameters of
interest. In practice, however, it may be hard to develop an algorithm that
ignores the corrupted entries in a statistically efficient way.

\subsection{Algorithm}
\label{ssec:algorithm}

The resulting algorithm is as follows:
\begin{enumerate}
\item {\it Initialization} Set the iteration counter $i = 1$ and initialize
  $\widehat{\bsigma}_n^{[0]}$ based on any prior information if available,
  otherwise initialize $\widehat{\bsigma}_n^{[0]}$ to zero. Initially use $\bW
  = \widehat{\bR}^{-1/2}$.
\item {\it Estimate} $\widehat{\btheta}^{[i]}$ by solving the WLS problem
  formulated in Eq. \eqref{eq:WLS_theta2} using $\widehat{\bsigma}_n^{[i-1]}$
  as prior knowledge.
\item {\it Estimate} $\widehat{\bsigma}_n^{[i]}$ using
  Eq. \eqref{eq:WLS_nuisance} using $\widehat{\btheta}^{[i]}$ as prior
  knowledge.
\item {\it Update} $\bW = \bR^{-1/2}$ to avoid the bias mentioned in the
  previous section.
\item {\it Check for convergence},
    otherwise continue with step 2.
\end{enumerate}

An algorithm that alternatingly optimizes for distinct groups of parameters,
in our case $\btheta$ and $\bsigma_n$, can be proven to converge if the value
of the cost function decreases in each iteration. We assume that a suitable
method is available to find $\btheta$. Since we propose to estimate
$\bsigma_n$ using the well known standard solution for least squares
estimation problems, the value of the cost function will decrease in both
steps, thus ensuring convergence. Although there is no guarantee that the
algorithm will converge to the global optimum, practical experience with LOFAR
and results from Monte Carlo simulations in this paper and in earlier papers
\cite{Wijnholds2006-1, Wijnholds2008-1} indicate that the proposed method
produces good results for most reasonable initial estimates.

\section{Experimental results}
\label{sec:results}

\subsection{The LOFAR prototype station}

\begin{figure}
\begin{minipage}{1.0\linewidth}
  \centering
  \centerline{\epsfig{figure=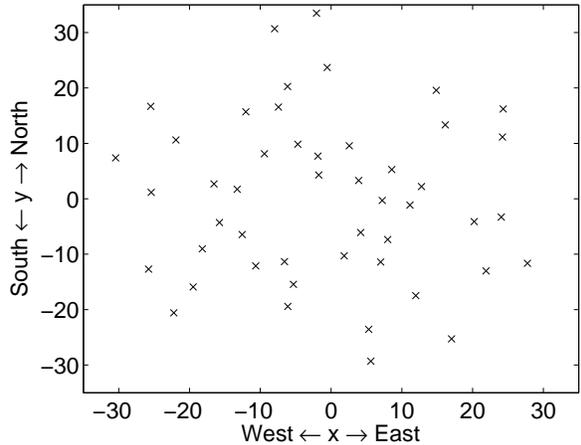,width=8.5cm}}
\end{minipage}
\caption{Array configuration of the 48 antenna LOFAR prototype
  station. \label{fig:CS10config}}
\end{figure}

The first full-scale LOFAR prototype station with real-time backend became
operational in the second quarter of 2006 \cite{Gunst2008-1}. This station
consisted of 48 dual-polarization antennas operating between 10 and 90 MHz
arranged in a randomized configuration based on rings with exponentially
increasing radii as shown in Fig.\ \ref{fig:CS10config}. Each of the two
signals from every antenna was filtered and digitized using a 12-bit 200 MHz
ADC. A real-time FPGA based digital processing backend splits the 100 MHz
Nyquist sampled base band in 512 subbands, each 195 kHz wide, using a
polyphase filter. The backend also provides a correlator which can correlate
in real-time the data from the 96 input channels for a single subband. This
subband may be any of the 512 available subbands and this choice may change
every second.

\subsection{Results}

For the demonstration in this paper we used data from a single 1 s snapshot
observation in the 195 kHz subband centered at 50 MHz. This observation was
done on 14 February 2008 at 1:42:07 UTC. We will calibrate the data using the
method described in Sec.\ \ref{ssec:algorithm}, where we will model correlated
noise terms on all baselines shorter than four wavelengths. The complete WLS
problem thus implies estimation of the amplitudes (48 parameters) and phases
(47 parameters) of the antenna based complex gains, the source power ratio of
the two brightest sources (1 parameter) and 764 real valued nuisance
parameters describing all non-zero entries of the noise covariance matrix for
a total of 860 free real valued parameters per polarization. In this
experiment we show that use of such nuisance parameters can reduce the complex
source structure on the sky to a simple model with just two point sources.

\begin{figure}
\begin{minipage}{1.0\linewidth}
  \centering
  \centerline{\epsfig{figure=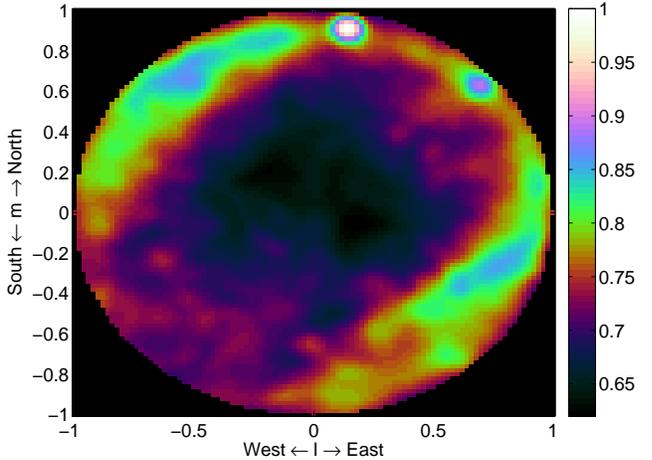,width=8.5cm}}
\end{minipage}
\caption{Calibrated all-sky map for a single polarization at 50 MHz from the
  48-element LOFAR prototype station. The image shows the sky projected on the
horizon plane of the station.}
\label{fig:skymapxcal}
\end{figure}

Figure \ref{fig:skymapxcal} shows a calibrated all sky map for a single
polarization. There are two bright point sources near the northeastern
horizon. The image also shows a lot of extended emission from the galactic
plane (on the northwestern horizon) and the north polar spur (on the eastern
horizon). This extended emission is hard to model accurately, but only affects
the short baselines since short distances in the aperture plane of a phased
array correspond to low spatial frequencies, which describe the structure on
large spatial scales.

\begin{figure}
\begin{minipage}{1.0\linewidth}
  \centering
  \centerline{\epsfig{figure=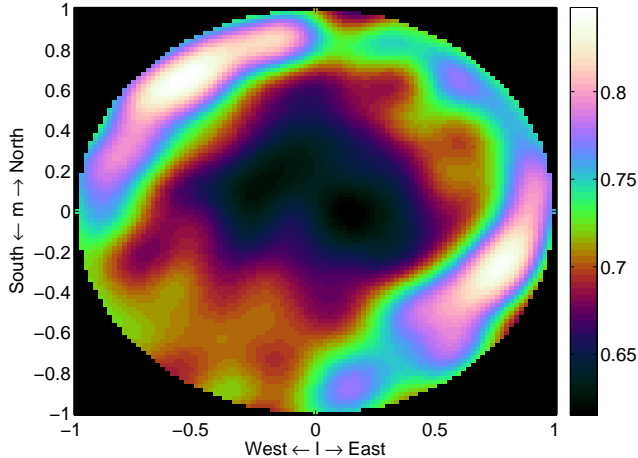,width=8.5cm}}
\end{minipage}
\caption{Calibrated all-sky map of extended emission observed at baselines
  shorter than four wavelengths for a single polarization at 50 MHz.}
\label{fig:skymapextxcal}
\end{figure}

It was found that most of this extended emission is captured by the
correlations on baselines shorter than four wavelengths. This affects 358
crosscorrelations and the 48 autocorrelations resulting in the aforementioned
764 real valued parameters to describe the non-zero entries of
$\bSigma_n$. Using the procedure outlined in the previous section,
$\widehat{\bsigma}_n$ was estimated simultaneously with $\widehat{\btheta}$
containing the other parameters. $\widehat{\bSigma}_n$ can therefore be
interpreted as an estimate of the extended source structure, noise coupling
and receiver noise powers. This is nicely demonstrated in Fig.\
\ref{fig:skymapextxcal} which shows an image based on $\widehat{\bSigma}_n$
after the calibration was completed.

\begin{figure}
\begin{minipage}{1.0\linewidth}
  \centering
  \centerline{\epsfig{figure=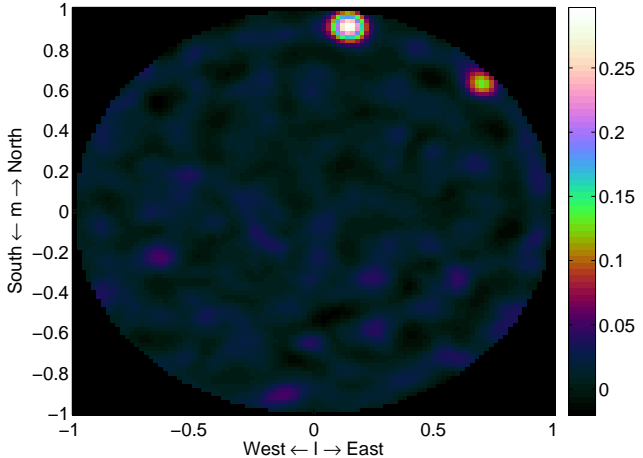,width=8.5cm}}
\end{minipage}
\caption{Difference between the calibrated all-sky map shown in
  Fig.\ \ref{fig:skymapxcal} and the contribution of extended emission shown in
Fig.\ \ref{fig:skymapextxcal} showing that the remainder can be accurately
modeled using a model with just two point sources.}
\label{fig:diffmap}
\end{figure}

Figure \ref{fig:diffmap} shows the difference between the maps shown in
Figs.\ \ref{fig:skymapxcal} and \ref{fig:skymapextxcal}. This shows that
$\bSigma_n$ provides a description of the extended emission that is
sufficiently accurate to reduce $\widehat{\bR} - \widehat{\bSigma}_n$ to an
array covariance matrix that can be described by a model consisting of only
two point sources. This thus reduces our original problem to one that has been
discussed extensively in the array signal processing literature.

\section{Improving the computational efficiency}

The calculation of $\widehat{\bsigma}_n$ using Eq.\ \ref{eq:WLS_nuisance}
forms the most expensive part of the algorithm in terms of CPU and memory
usage due to the Kronecker products. These Kronecker products can only be
reduced to simpler Khatri-Rao or Hadamard products in a number of special
cases, such as a diagonal noise covariance matrix treated in
\cite{Wijnholds2008-1}. However, the parameterization of $\bSigma_n$ chosen
here implies that each entry of the array covariance matrix with a
contribution from $\bSigma_n$ is affected by a unique additive
parameter. Intuitively, one would therefore expect that the weighting in Eq.\
\eqref{eq:WLS_nuisance} would not make much difference. Omitting this would
reduce the CPU and memory requirements considerably, since the Kronecker
products and the inverse increase the numerical complexity from $o \left ( N
  P^2 \right )$ to $o \left ( N^3 + P^6 \right )$ and the size of the largest
matrix from $P^2 \times N$ to $P^2 \times P^2$, where $N$ is the number of
noise parameters stacked in $\bsigma_n$.

\begin{table}
\renewcommand{\arraystretch}{1.3}
\begin{minipage}[b]{1.0\linewidth}
\centering
\begin{tabular}{l|rrr}
\hline
\hline
\# & $l$ & $m$ & $\sigma_q^2$\\
\hline
1 & 0.24651 & -0.71637 & 1.00000\\
2 & -0.34346 & 0.76883 & 0.88051\\
3 & -0.13125 & -0.31463 & 0.79079\\
4 & -0.29941 & -0.52339 & 0.74654\\
5 & 0.39290 & 0.58902 & 0.69781\\
\hline
\hline
\end{tabular}
\end{minipage}
\caption{Source powers and source locations used in the simulations}
\label{tab:srcmodel}
\end{table}

This idea was therefore tested in Monte Carlo simulations. For these
simulations, a five armed array was defined, each arm being an eight-element,
one wavelength spaced ULA. The first element of each arm formed an equally
spaced circular array with half wavelength spacing between the elements. The
source model is presented in Table \ref{tab:srcmodel}. This source model was
generated with a random number generator to verify that the proposed approach
works for arbitrary source models.

\begin{figure}
\begin{minipage}{1.0\linewidth}
\centering
  \centerline{\epsfig{figure=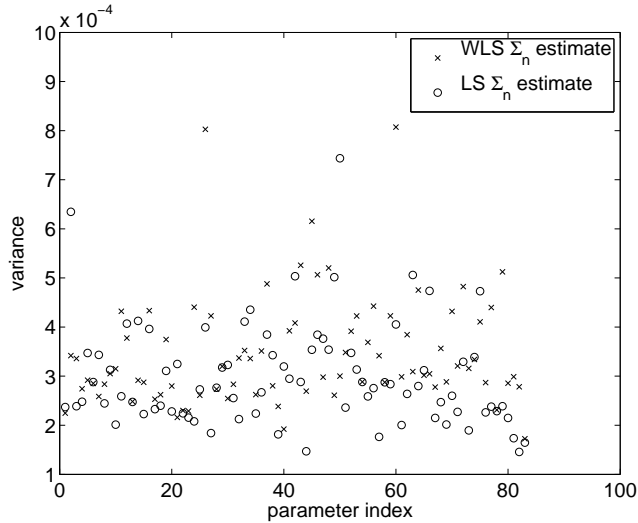,width=8.5cm}}
\end{minipage}
\caption{Comparison of the variance on the direction independent complex gain
  and the source power estimates obtained in Monte Carlo simulations with
  weighted LS estimation of $\widehat{\bsigma}_n$ and unweighted LS estimation
  of $\widehat{\bsigma}_n$.}
\label{fig:MCsimres}
\end{figure}

Figure \ref{fig:MCsimres} compares the variance on the estimates for the
omnidirectional complex gains of the receiving elements and the source powers
obtained after 100 runs of a Monte Carlo simulation with weighted LS
estimation of $\widehat{\bsigma}_n$ and the computationally more efficient
unweighted LS estimation of $\widehat{\bsigma}_n$. The complex receiving
element gains and the source powers, stacked in $\btheta$, were estimated
using weighted least squares in both cases. Both algorithms generally
converged within three iterations to relative error per parameter of $\sim
10^{-4}$, i.e.\ well below the Cram\`er-Rao bound. The convergence rate in
these simulations was one digit per iteration down to the numerical accuracy
provided by double precision floating point numbers.

The results indicate that the variance on these estimates is the same in both
cases within the accuracy provided by the simulations. We therefore conclude
that it is viable to discard the weighting in Eq.\
\eqref{eq:WLS_nuisance}. With this modification all 860 free parameters in the
experiment described in the previous section could be extracted from the
actual data using Matlab running on a standard dual core 2.4 GHz CPU in only
0.4 seconds. This implies that a single 2.4 GHz core can keep up with the data
from the correlator at the LOFAR station, which real-time correlates the
antenna signals for a single subband with one second integration time. This
update rate is required to track variations in the electronic gains and the ionosphere.

\section{Conclusions}

We have demonstrated using data from a LOFAR prototype station that the
effects of noise coupling, receiver noise powers and extended emission on a
radio astronomical phased array can be phenomenologically described by a
non-diagonal noise covariance matrix with non-zero entries on short
baselines. These entries can be computationally efficient and accurately
estimated by a WALS algorithm alternating between estimation of the correlated
noise parameters and calibration parameters.

\bibliographystyle{IEEEbib}
\bibliography{refs}

\end{document}